\documentclass[preprint,3p,authoryear]{elsarticle}
\usepackage{graphicx}
\usepackage{amssymb}
\usepackage{lineno}
\usepackage{tabularx}
\begin{document}
\begin{frontmatter}
\title{DEVELOPMENT OF MODELS OF ACOUSTIC-GRAVITY WAVES  IN THE UPPER ATMOSPHERE (OVERVIEW)}
\author{O. K. Cheremnykh}
\ead{oleg.cheremnykh@gmail.com}
\author{A. K. Fedorenko}
\ead{fedorenkoak@gmail.com}
\author{E. I. Kryuchkov}
\ead{kryuchkov.ye@gmail.com}
\author{Yu. O. Klymenko$^*$}
\ead{yurklym@gmail.com}
\cortext[cor1]{Corresponding author.}
\author{I. T. Zhuk}
\ead{zhukigor@gmail.com}

\address{Space Research Institute of  the National Academy of Sciences of Ukraine and the State Space Agency of Ukraine, prosp. Akad. Glushkova 40, build. 4/1, 03187, Kyiv, Ukraine}

\begin{abstract}
We present the results of studies of acoustic-gravity waves (AGWs) in the upper atmosphere of the Earth. The work has been mainly aimed at studying the peculiarities of atmospheric AGWs in the atmosphere based on the theoretical models taking into account the properties of the real atmosphere and the model verification based on spacecraft measurement data.
This overview includes the main results obtained by the authors over the last years. The details of the calculations are contained in the authors' works cited here.

The possibility of the existence of new types of evanescent acoustic-gravity waves in the isothermal atmosphere is theoretically predicted. A previously unknown inelastic ($\gamma $-) mode and a family of evanescent pseudo-modes are discussed. The possibility of observing evanescent modes on the Sun and in the Earth's atmosphere is analyzed. The peculiarities of the propagation of the acoustic-gravity waves at the boundary of two isothermal half-spaces with different temperatures are studied in dependence of their spectral parameters and the magnitude of the temperature jump at the boundary. It is shown that such waves may be observed in the atmosphere where the temperature changes sharply with the height, for example, in the lower part of the Earth's thermosphere or in the chromosphere-corona transition region on the Sun.

The peculiarities of the interaction of acoustic-gravity waves with spatially inhomogeneous atmospheric flows are also studied. It is analyzed the observed effects that are a consequence of such interaction. It is highlighted the azimuths of the wave propagation, the change in their amplitudes, and the blocking effect in the counterflow. The influence of vertical non-isothermality on propagation of acoustic-gravity waves including the modification of acoustic and gravitational regions depending on the temperature is studied.

On the basis of modified Navier-Stokes and heat transfer equations, the influence of attenuation on the propagation of acoustic-gravity waves in the atmosphere is analyzed. For the first time, in addition to usually considered velocity gradient, we take into account the additional wave transfer of momentum and energy due to density gradient. It is also considered the attenuation of various types of evanescent acoustic-gravity waves in the atmosphere.

It is shown that the rotation of the atmosphere leads to the modification of the continuous spectrum of evanescent acoustic-gravity waves with frequencies greater than the Coriolis parameter. On the diagnostic diagram, this spectrum fills the entire "forbidden" region between the freely propagating acoustic waves and the internal gravitational ones.
\end{abstract}
\end{frontmatter}

\section{INTRODUCTION}

Acoustic-gravity waves (AGWs) have been intensively studied in the physics of the Earth's atmosphere and the Sun for more than 60 years. Space missions also have been served as an additional stimulus to study these waves in atmospheres of other planets, for example, Mars and Venus [Forbes, Moudden, 2009; Schubert, Walterscheid, 1984]. To a great extent, the interest to AGWs is related to the important role of these waves in dynamics and energetics of planetary atmospheres and the Sun, providing an effective redistribution of disturbance energy on a global scale.

In the Earth's atmosphere, these waves can be generated by various sources of natural and anthropogenic origins. AGWs are associated with "influences from above" if their sources are localized in the upper atmosphere. It is, for example, precipitation of charged particles in high latitudes, atmospheric flows, movement of the solar terminator, etc. [Somsikov, 1983; Bespalova  et al, 2016; Hajkovicz, 1991; Fritts, Vadas, 2008; Prikryl et al., 2005]. Considerable attention is now being paid to ionospheric study of the waves caused by tropospheric or ground sources. It is so-called effects of the influences on the ionosphere and the atmosphere "from below" [Rapoport et al., 2004; Sauli, Boska, 2001; Yi$\rm\check{g}$it et al., 2008].

When AGWs propagate in the atmosphere from the bottom up, their amplitudes rapidly increase with the height due to the decrease of the background density [Hines, 1960; Yeh, Liu, 1974]. In this regard, when considering them, it is often necessary to take into account nonlinear effects. In recent years, significant progress has been made in the development of the nonlinear theory of AGW [Stenflo, Shukla, 2009; Huang et al., 2014]. The main difficulties of the analytical approach in studying the atmosphere are that its properties (density, temperature, transport coefficients) change with the height. As a result, the system of hydrodynamic equations describing the wave disturbances in the atmosphere contains height-dependent coefficients. For an isothermal atmosphere, the barometric distribution of the density with the height is usually excluded in the AGW theory using a well-known substitution [Hines, 1960; Yeh, Liu, 1974], which leads to a system of equations with constant coefficients. However, it is impossible to take into account the changes with temperature and transfer coefficients in a similar way. Therefore, an analytical expression for the wave dispersion in a viscous non-isothermal atmosphere can be obtained only in a local approximation, assuming that the parameters of the medium change rather slowly on the wavelength scales. The local dispersion equation of the AGW was obtained and analyzed in [Vadas et al., 2005; Fedorenko et al., 2021] taking into account the viscosity and the thermal conductivity. Numerical modeling of AGWs in a realistic atmosphere, including the effects of viscosity and thermal conductivity, is an important direction of modern researches of these waves [Zhang, Yi, 2002; Cheremnykh et al., 2010].

The linear theory of AGW allows the existence of a continuous spectrum of freely propagating waves in the atmosphere, consisting of acoustic and gravitational regions, as well as a family of horizontally propagating evanescent wave modes. In particular, the evanescent waves include the well-known Brunt-V$\rm\ddot{a}$iss$\rm\ddot{a}$l$\rm\ddot{a}$ (BV) oscillations, the Lamb wave, the non-divergence \textit{f}-mode [Rosental, Gough, 1994; Ghosh  et al., 1995], as well as the recently discovered inelastic $\gamma $-mode [Cheremnykh et al., 2019]. Evanescent waves are effectively generated in those regions of the atmosphere where there are significant vertical gradients of temperature or density, for example, at the boundary between the chromosphere and the corona on the Sun [Rosental, Gough, 1994]. In the Earth's atmosphere, such waves may occur in the lower part of the thermosphere or at the heights of the tropopause and the mesopause.

This article presents the resent author research results on atmospheric acoustic-gravity wave, taking into account the characteristics of the real atmosphere. In the first section, the evanescent wave modes in the atmosphere are investigated and their differences from freely propagating AGWs are indicated. Section 2 considers the propagation of acoustic-gravity waves at the boundary of two isothermal media having different temperatures, as well as spectrum modification and polarization properties of waves at the boundary. In the third section, the effect of temperature change with height on the propagation of AGWs is studied, as well as the modification of the acoustic and gravitational regions of the spectrum is investigated. In section 4, the possibility of realizing of evanescent acoustic-gravity modes in a non-isothermal atmosphere with a continuous altitudinal temperature profile is studied. The fifth section presents the results of research on the interaction of AGWs with horizontally inhomogeneous flows. In section 6 we investigate the attenuation of various types of acoustic-gravity waves based on the modified Navier-Stokes and heat transfer equations. The seventh section shows how the rotation of the atmosphere affects the propagation and modification of the spectrum of evanescent waves.

\section{EVANESCENT WAVE MODES IN ISOTHERMAL ATMOSPHERE}

The most famous among evanescent waves are the horizontal Lamb wave and vertical oscillations of Brunt-V$\rm\ddot{a}$iss$\rm\ddot{a}$l$\rm\ddot{a}$ [Waltercheid, Hecht, 2003]. In hydrodynamics and in physics of atmosphere and the Sun, it is also known the \textit{f}-mode with dispersion $\omega ^{2} =k_{x} g$ [Tolstoy, 1963; Jones, 1969]. Experimental observations of the \textit{f}-mode on the Sun are used to diagnose flows, specifying the dimensions of the solar radius and for other features of the structure and dynamics of the Sun [Rosental, Gough, 1994; Ghosh et al., 1995]. In the Earth's atmosphere, evanescent waves are observed at altitudes near the mesopause [Simkhada et al., 2009].

We investigated the peculiarities of the propagation of evanescent wave modes in the atmosphere and theoretically predicted their new types ($\gamma $-mode and a family of pseudo-modes). Note that the discovery of the $\gamma $-mode gives the possibility to combine the evanescent acoustic-gravity modes of the isothermal atmosphere into the entired system.

It is shown that each evanescent mode can be associated with a pseudo-mode. Both of them coincides in the dispersion, but differs in the polarization and the height amplitude dependence [Cheremnykh et al., 2019]. The properties of the main evanescent modes for the isothermal atmosphere are summarized in Table 1 (according to [Cheremnykh et al., 2019]).  In the Table 1: $V_{x} $ and $V_{z} $ are the wave perturbations of the horizontal and vertical components of particle velocity $\sim\exp(az)\exp[i(\omega t-k_x x)]$, $a$ is the height altitude dependence, $\omega $ is the wave frequency, $k_{x} $ is the horizontal component of the wave vector, $c_{s} =\sqrt{\gamma gH} $ is the sound speed,  $N=\sqrt{g\left(\gamma -1\right)/\gamma H} $ is Brunt-V$\rm\ddot{a}$iss$\rm\ddot{a}$l$\rm\ddot{a}$ oscillation, $H$ is the atmosphere scale height, $g$ is the acceleration of the gravity, and $\gamma $ is the ratio of specific heats.

On the spectral $\omega \left(k_{x} \right)$-plane, the location of the dispersion curves of the main evanescent modes in the atmosphere is shown in Fig. 1 relative to the gravitational and acoustic regions of freely propagating AGWs [Cheremnykh et al., 2019]. The dispersion dependence $\omega ^{2} =k_{x} g\left(\gamma -1\right)$ corresponding to the $\gamma $- and $\gamma _{p} $- modes touches the gravitational region of freely propagating AGWs at the same value $k_{x} =1/2H$ at which the curve $\omega ^{2} =k_{x} g$ (dispersion of $f$- and $f_p$-modes) touches the acoustic region (Fig. 1). The dispersion curves $\omega ^{2} =k_{x} g$ and $\omega ^{2} =k_{x} g\left(\gamma -1\right)$ are symmetric with respect to the "characteristic" Beer curve [Beer, 1974], which separates the acoustic region from the gravitational one.

\begin{figure}
\centering
\includegraphics[scale=1.0]{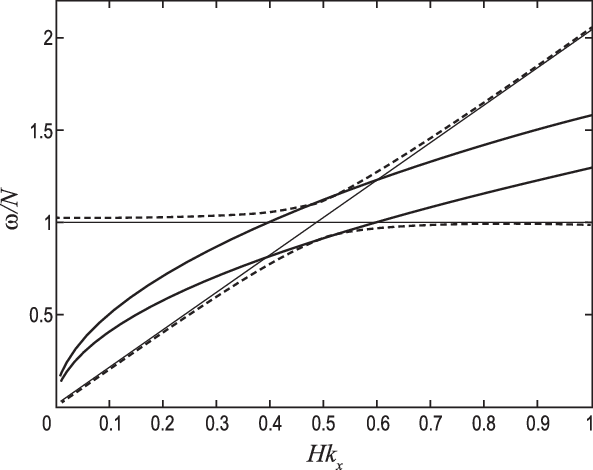}
\caption{Dispersion dependencies of evanescent wave modes: $\omega =\sqrt{k_{x} g} $ and  $\omega =\sqrt{k_{x} g\left(\gamma -1\right)} $ (upper and lower solid curves, respectively), $\omega =N$ (horizontal line), and $\omega =k_{x} c_{s} $ (sloping line) [Cheremnykh et al., 2019]. The boundaries of  acoustic and gravitational regions of freely propagating AGWs are shown by dashed curves.}
\end{figure}

Table 1. Main types of evanescent acoustic-gravity wave modes in isothermal atmosphere [Cheremnykh et al., 2019].
\begin{center}
\begin{tabular}{|p{1.5in}|p{1.1in}|p{0.65in}|p{2.1in}|} \hline
Type of mode & Dispersion & Height altitude dependence   $a$& Polarization \\ \hline
Lamb wave  & $\omega ^{2} =k_{x}^{2} c_{s}^{2}$ & $\frac{\gamma -1}{\gamma H} $ & $V_{z} =0$; $V_{x} \ne 0$ \\ \hline
Lamb's pseudo-mode & $\omega ^{2} =k_{x}^{2} c_{s}^{2}$ & $\frac{1}{\gamma H} $ & $V_{x} \left(2-\gamma \right)k_{x} g=i\left(N^{2} -k_{x}^{2} c_{s}^{2} \right)V_{z} $ \\ \hline
BV oscillation  & $\omega ^{2} =N^{2} $  & $\frac{1}{\gamma H} $ & $V_{x} =0$; $V_{z} \ne 0$ \\ \hline
BV pseudo-mode &$\omega ^{2} =N^{2}$  & $\frac{\gamma -1}{\gamma H} $ & $V_{x} \left(k_{x}^{2} c_{s}^{2} -N^{2} \right)=i\left(2-\gamma \right)k_{x} gV_{z} $ \\ \hline
Non-divergence mode (\textit{f}-mode), $div\vec{V}=0$ &  $\omega ^{2} =k_{x} g$ & $k_{x} $ & $V_{x} =-iV_{z} $ \\ \hline
Pseudo non-divergence mode ($f_p$-mode)  & $\omega ^{2} =k_{x} g$ & $\frac{1}{H} -k_{x} $ & $V_{x} \left(\frac{1}{\gamma H} -k_{x} \right)=-i\left(k_{x} -\frac{\gamma -1}{\gamma H} \right)V_{z} $ \\ \hline
Inelastic mode ($\gamma $-mode), $div\left(\rho \vec{V}\right)=0$ &  $\omega ^{2} =k_{x} g\left(\gamma -1\right)$ & $\frac{1}{H} -k_{x} $ & $V_{x} =iV_{z} $ \\ \hline
Pseudo inelastic mode ($\gamma _{p} $-mode) & $\omega ^{2} =k_{x} g\left(\gamma -1\right)$ & $k_{x} $ & $V_{x} \left(k_{x} -\frac{\gamma -1}{\gamma H} \right)=i\left(\frac{1}{\gamma H} -k_{x} \right)V_{z} $ \\ \hline
\end{tabular}
\end{center}

\section{WAVES AT THE BOUNDARY OF TWO ISOTHERMAL MEDIA}

For horizontal waves in the model of an infinite atmosphere, it is not possible to ensure simultaneously the energy decays up and down from the level of the wave propagation. The simplest model of the atmosphere, where such a condition can be fulfilled, is two isothermal half-spaces with different temperatures. The properties of evanescent acoustic-gravity waves propagating at the boundary of two media with different temperatures were analyzed in detail in works [Cheremnykh et al., 2019; Fedorenko et al., 2022].

The vertical velocity components $V_{z} $ and the Lagrangian pressure $\Psi =\frac{\partial p'}{\partial t} +V_{z} \frac{\partial p}{\partial z} $  must be continuous at the boundary of two media [Tolstoy, 1963; Rosental, Gough, 1994]. It comes to kinematic ($V_{z1} =V_{z2} $) and dynamic ($\Psi _{1} =\Psi _{2} $) boundary conditions, respectively. In addition to these boundary conditions, the wave energy density $E\sim \rho \left(z\right)\left(V_{x}^{2} +V_{z}^{2} \right)$ must decreases up and down from the boundary interface. Finaly, it leads to the dispersion dependencies presented in Fig. 2. Here $T_{1} $ and  $T_{2} $ are the temperatures in the lower and upper half-spaces, respectively. Such dependencies were first obtained in [Cheremnykh et al., 2019].

\begin{figure}
\centering
\includegraphics[scale=0.8]{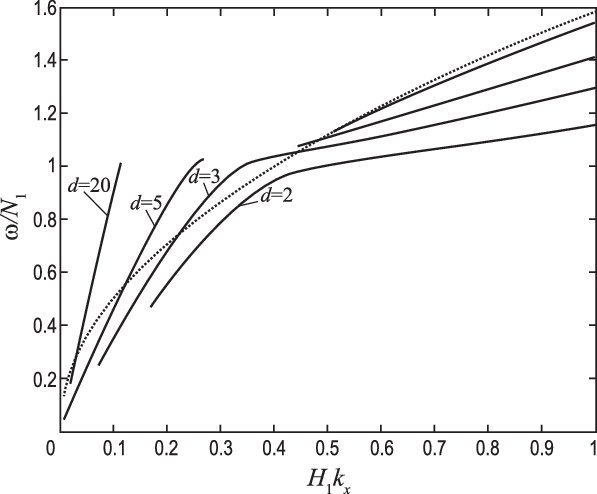}
\caption{Dispersion dependencies $\omega =f\left(k_{x} \right)$ at the boundary of two media with different values of parameter $d=T_{2} /T_{1} $. Here $N_1$ and $H_1$ are  the Brunt-V$\rm\ddot{a}$iss$\rm\ddot{a}$l$\rm\ddot{a}$ frequency and atmospere scale height for media 1. The dotted curve shows the dependence $\omega ^{2} =k_{x} g$ [Cheremnykh et al., 2019].}
\end{figure}

The matching of evanescent solutions at the boundary leads to a modification of the properties of evanescent waves in comparison with the solutions obtained for an infinite atmosphere. A notable feature of the dispersion curves in Fig. 2 is the cut-offs, whose locations in the $\omega \left(k_{x} \right)$ spectral plane coincides with the boundaries for free wave propagation regions. The reason of these cut-offs is due to the fact that the temperatures in the lower and upper half-spaces are different. Therefore, the evanescent regions in the two half-spaces do not coincide. Waves propagating at the boundary exist only in the overlap region of these two evanescent regions. As a result, in certain regions of the spectral diagram we have cut-offs of the dispersion curves.

At large values of $d=T_{2} /T_{1} $ and for $k_{x} H_{1} >0.5$ ($H_1$ is the atmospheric scale height for media 1), the dispersion dependence of the boundary waves approaches the pure \textit{f}-mode dispersion $\omega ^{2} =k_{x} g$ (dashed curve). At large $d$ and small values of $k_{x} $, the dispersion dependencies are close to linear, which is characteristic of acoustic-type waves. At $d<4$ the dispersion curve is continuous with the bottom cut-off at the border of the free propagation region (Fig. 2). It means that at different values of $k_{x} $, the boundary condition and the requirement of the energy reduction are fulfilled simultaneously along the entire length of the dispersion curve. However, at $d>4$ the spectrum of the waves at the boundary of the medium splits into two different branches, separated by the discontinuity region. In the spectral range corresponding to the discontinuity, the waves satisfy the boundary condition, but they cannot be simultaneously decreasing in energy up and down from the border [Cheremnykh et al., 2019; Fedorenko et al., 2022]. The width of this energetically "forbidden" region increases with increasing parameter $d$. Note that the behavior of the dispersion curve of the boundary waves at small values of $k_{x} $ is noticeably different from the pure \textit{f}-mode dispersion. In this case, at the boundary it is realized the acoustic-type waves with close to linear dependence $\omega \left(k_{x} \right)$.

For the lower part of the Earth's thermosphere, it can be accepted $d=T_{2} /T_{1} \approx 3\div 4$ at average solar activity. Therefore, in the temperature discontinuity model for the entire range of $k_{x} $, there can be waves that simultaneously satisfy the boundary conditions and the condition of decreasing energy up and down from the boundary. For the transition region of the chromosphere-corona on the Sun we have $d\approx 40$. For the \textit{f}-mode observed on the Sun, no discontinuities in the dispersion dependences were observed. For the conditions of the Sun's atmosphere, more complex models should be used. They have to take into account the finite dimensions of the transition region, the influence of the magnetic field, the presence of plasma, and other real features.

\section{INFLUENCE OF VERTICAL NON-ISOTERMITY ON AGW PROPAGATION }

In the Earth's atmosphere, there are altitude intervals where the temperature and chemical composition change rather slowly. The propagation of waves at these altitudes can be described within the framework of the linear theory of AGWs developed for an isothermal atmosphere. However, in the general case, the altitude profile of the temperature in the atmospheres of the planets and the Sun is a complex function, which is determined by the balance between the energy input and the output mechanisms existing in the given altitude region. As a result, at some altitudes in the atmosphere, there are regions with a sharp altitudinal temperature gradient, where maxima and minima are formed. In such the regions, the AGW theory of an isothermal atmosphere becomes unacceptable. The complex altitude dependence $T\left(z\right)$ determines the specifics of the acoustic-gravity wave propagation at different altitude levels of the atmosphere, which, among other things, can lead to their reflection or waveguide propagation.

Modern numerical models make it possible to take into account a change in the temperature with the height, the viscosity, the thermal conductivity, the wind, the ionic friction and other features of the real atmosphere, which lead to deviations from the idealized theory of AGWs [Cheremnykh et al., 2010; Zhang, Yi, 2002]. However, it is important to understand how the dispersion of AGW is modified under conditions of a non-isothermal atmosphere. In the work [Fedorenko et al., 2020a], it was obtained an analytical expression for the dispersion of AGW in the atmosphere, if the temperature changes slowly with the height. From this dispersion, it follows that at fixed values of $\omega $ and $k_{x} $, the value of the vertical component of the wave vector changes with the height as the temperature changes, i.e., the wave undergoes the refraction. If for the wave with certain values  of $\omega $ and $k_{x} $, there is inequality $k_{z}^{2} >0$ in a certain height interval, then we have $k_{z}^{2} <0$ above and below this interval and here it is formed an atmospheric waveguide.

The isothermal configuration of the acoustic (Ac) and gravitational (Gr) regions of free propagation AGWs with $k_{z}^{2} >0$ is shown in Fig. 3a at $dH/dz=0$ [Fedorenko et al., 2020a]. When the temperature changes, these areas are modified, as shown in Figs. 3b, 3c and 3d. In a non-isothermal atmosphere, the location of free ($k_{z}^{2} >0$) and evanescent ($k_{z}^{2} <0$) regions on the spectral $\left(\omega ,k_{x} \right)$-plane changes depending on the magnitude and the sign of $dH/dz$. At $dH/dz<0$, Ac- and Gr-regions diverge, while the size of the "forbidden" region increases (Fig. 3b). At the value of $dH/dz=-\left(\gamma -1\right)/\gamma $ when $N^{2} =0$, the gravitational branch disappears and the atmosphere becomes convectively unstable.

\begin{figure}
\centering
\includegraphics[scale=0.6]{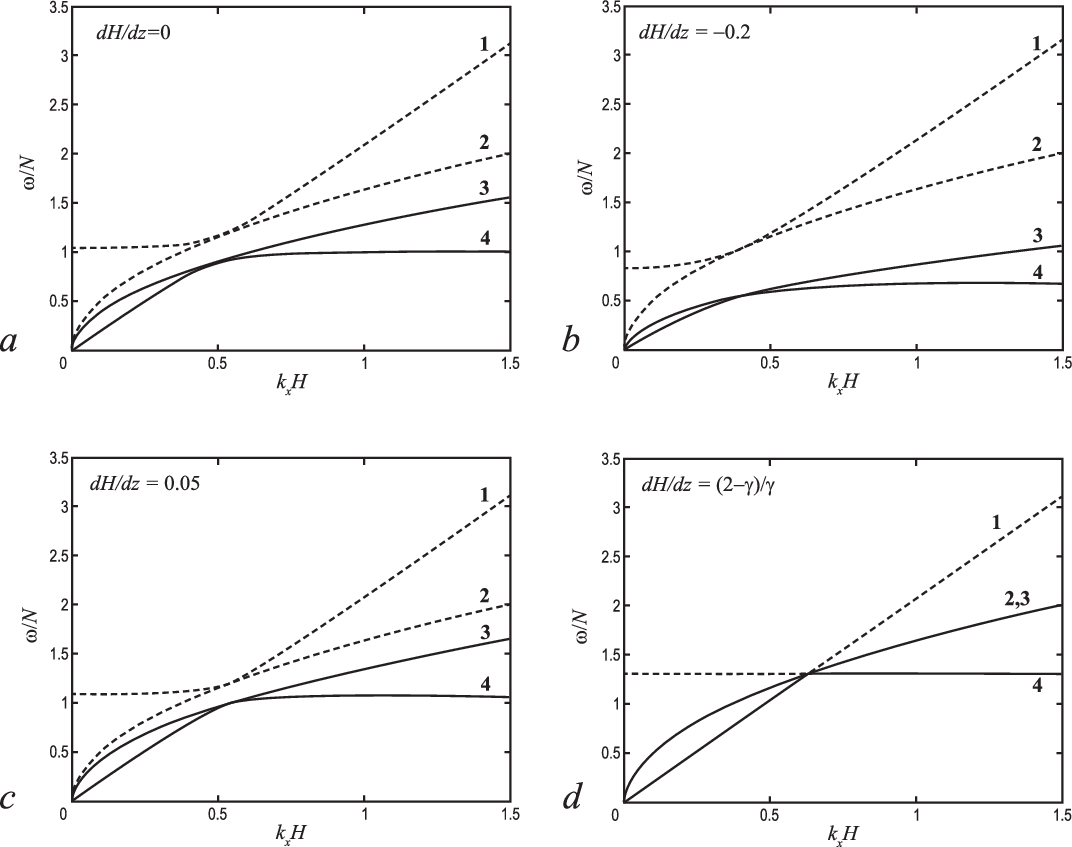}
\caption{Dispersion dependences of AGW at different values of $dH/dz$. Dashed lines show the boundaries of the acoustic region of freely propagation AGWs (curve 1) and the non-divergent mode $\omega =\sqrt{k_{x} g} $ (curve 2). Solid lines indicate the dispersion of the inelastic mode $\omega =\sqrt{k_{x} g\left(\gamma -1+\gamma dH/dz\right)} $ (curve 3) and the boundaries of the gravitational region of freely propagating AGWs (curve 4) [Fedorenko et al., 2020a].}
\end{figure}


With the increase of $dH/dz>0$, the width of the evanescent region decreases (Fig. 3c). At $dH/dz=\left(2-\gamma \right)/\gamma $, the acoustic and the gravitational regions close at point $k_{x}^{*} =1/\gamma H$ (see Fig. 3d). In the Earth's atmosphere with average solar activity, this condition is realized at altitudes of about 110-130 km.

\section{EVANESCENT WAVE MODES IN NON-ISOTHERMAL ATMOSPHERE}

The possibility of the existence of evanescent modes in a non-isothermal atmosphere with a continuous altitudinal temperature profile was investigated in [Cheremnykh et al., 2021a]. It is shown that in a non-isothermal atmosphere can be realized two modes: the \textit{f}-mode with the dispersion coinciding with the isothermal case, and the $\gamma $\textit{-}mode with the dispersion modified by the height temperature changes. The condition of energy density decreasing up and down from propagation level is fulfilled for the \textit{f}-mode at the heights of the local temperature minima. A similar energy condition for the existence of $\gamma $-modes is fulfilled at the heights of the local temperature maxima. Accordingly, in the Earth's atmosphere, it is possible to distinguish the possible altitude intervals of the existence of these modes, see Fig. 4. The area of possible implementation of these modes is considered more broadly in Table 2.

\begin{figure}
\centering
\includegraphics[scale=0.8]{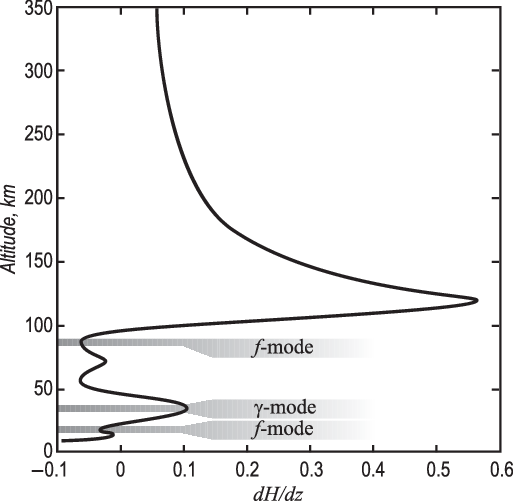}
\caption{Altitude dependence of $dH/dz$ and hypothetical areas of existence of \textit{f}- and $\gamma $-modes at altitudes of local extremes of temperature ($dH/dz=0$) in the Earth's atmosphere [Cheremnykh et al., 2021a].}
\end{figure}
Table 2. Implementation of evanescent modes in the atmosphere [Cheremnykh et al., 2021a].
\begin{center}
\begin{tabular}{|p{0.7in}|p{1.5in}|p{1.5in}|p{1.7in}|} \hline
Mode type & Dispersion equation & Height dependence amplitude  & Area of realization \\ \hline
\textit{f-}mode \newline  & $\omega ^{2} =k_{x} g$ & $a=k_{x} $ & The thermosphere, regions of $T\left(z\right)_{\min } $ in planet atmospheres, the mesopause and the tropopause of the Earth's atmosphere, the chromosphere of the Sun \\ \hline
$\gamma $\textbf{-}mode  & $\omega ^{2} =k_{x} g\left(\gamma -1+\gamma \frac{dH}{dz} \right)$ & $a=\frac{{\rm 1}}{H} \left(1+\frac{dH}{dz} \right)-k_{x} $ & The thermosphere, regions of $T\left(z\right)_{\max } $in planet atmospheres, the stratopause of the Earth's atmosphere \\ \hline
\end{tabular}
\end{center}

For the \textit{f}-mode, the condition of decreasing energy up and down from the propagation level is fulfilled at the heights of local minima $T\left(z\right)_{\min } $. At the same time, the characteristic wavelength of the \textit{f}-mode is $\lambda _{x} =4\pi H_{0} $ ($H_0$ is the atmospheric scale hight in the min- or max- temperature regions). These disturbances can be observed, for example, in the mesopause of the Earth with characteristic values of $\lambda _{x} \approx 75$km and $T_{0} \approx 235^{\circ } $C. In the chromosphere of the Sun, the spectral characteristics of the \textit{f}-mode are $\lambda _{x} \approx 1600$km and $T_{0} \approx 194^{\circ } $C. In the regions of maximum temperature $T\left(z\right)_{\max } $, for example, at the height of the Earth's stratopause, the $\gamma $-mode may be realized. Moreover, the period of this mode is slightly longer than the period of Brunt-V$\rm\ddot{a}$iss$\rm\ddot{a}$l$\rm\ddot{a}$ at the height of its propagation, and the wavelength is $\lambda _{x} =4\pi H_{0} \approx 100$km.

\section{PROPAGATION OF AGW IN SPATIALLY NON-HOMOGENEOUS FLOWS}

Observations on the Dynamic Explorer 2 (DE2) satellite  show the predominance of AGWs with certain spectral properties in the polar thermosphere. Namely, the horizontal wave lengths contain $\sim$500--700 km and the frequencies are closed to the BV frequency. Also, according to satellite data for the polar regions, AGWs systematically propagate towards the wind, and their amplitudes increase with increasing flow speed [Fedorenko, Kryuchkov, 2013; Fedorenko et al., 2015]. These experimental results indicate the need to take into account the interaction of AGWs with the horizontally inhomogeneous flows.

If the flow is spatially uniform (i.e., if its speed does not depend on the coordinates), then it usually goes to the reference system connected with the moving medium. In the reference system, it is valid the theoretical relations of AGW (dispersion equation, polarization relations, etc.) obtained for the stationary medium. The same approach can be applied if the current velocity varies slowly on the wavelength scale. In the local approximation, the flow inhomogeneity practically does not affect the dispersion of AGW, but only the wave amplitudes [Lighthill, 1978].

In the local approximation, the modification of the gravitational branch of the AGW spectrum in the counterflow was investigated in [Fedorenko et al., 2018]. Some results of these studies are shown in Fig. 5. On each of the curves given, the condition $k_{z}^{2} =0$ is fulfilled for a fixed value of flow velocity $W_{x} =const$. The ``forbidden'' area ($k_{z}^{2} <0$) is located above, and the area of free propagation of AGWs ($k_{z}^{2} >0$) is below. The upper curve in Fig. 5 corresponds to the condition $W_{x} =0$ and coincides with the gravitational branch for the stationary medium. It can be seen that the region of free wave propagation shifts towards low frequencies for the stationary observer with the increase of $W_{x} $. AGWs with frequencies $\omega _{0} =0.5N$ (period $\approx$20 min), $0.3N$ ($\approx$35 min) and $0.17N$ ($\approx$60 min) cannot propagate towards the wind whose speed exceeds 300 m/s, 500 m/s and 600 m/s, respectively. Here, $\omega _{0} $ is the frequency measured by a stationary observer. Against particularly strong wind with a speed of more than $\sim$600 m/s, which is reached at geomagnitically disturbed polar thermosphere, only low-frequency AGWs can propagate with frequencies $<$0.001 s$^{-1}$.

\begin{figure}
\centering
\includegraphics[scale=0.8]{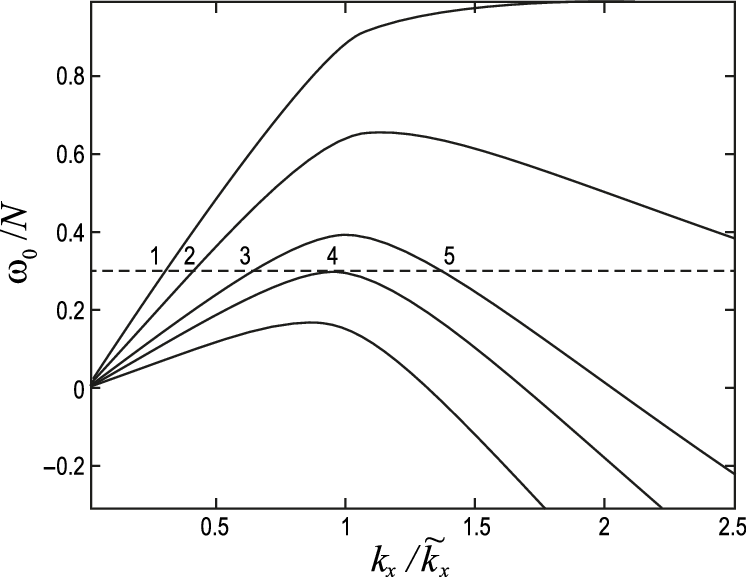}
\caption{Modification of gravity region of AGWs in horizontal flow. Frequency $\omega _{0} /N$ (in the reference system of stationary observer) is shown on the vertical axis and component of wave vector $k_{x} /\tilde{k}_{x} $, where $\tilde{k}_{x} =N/c_{s} $, is plotted on the horizontal axis. The curves correspond to different values of wind speed, $W_{x} $=0, --200, --400, --480, --600 m/s are plotted from top to bottom [Fedorenko et al., 2018].}
\end{figure}

An important feature of the AGW dispersion curves propagating towards the flow is the presence of maxima corresponding to the zero of the horizontal group velocity (see Fig. 5). Since there is no horizontal transfer of the wave energy at the maximum points, the wave amplitudes at these points can increase significantly. These points are called blocking ones for waves on the opposite current. The velocity of the blocking flow depends on the spectral parameters of AGWs. The limiting parameters of AGW blocking are calculated in [Fedorenko et al., 2018] for different values of fixed frequencies $\omega _{0} $.

The evolution of the wave with a frequency $\omega _{0} =0.3N$ that moves opposite to the increasing flow is shown in the Fig.5. Allowable range of $k_{x} $ in the non-moving medium is located to the right side of the point 1 (Fig.5). The value of $k_{x} $ shifts from point 2 to the right side when the flow velocity begins to increase. As the flow velocity increases further, the admissible range of $k_{x} $ is bounded on both sides (see the points 3 and 5) and it is collapsed to specific value of $\tilde{k}_{x} \sim N/c_{s} $ at some wind velocity $W_{x}^{*} $ (point 4). The value of AGW horizontal group velocity in this point becomes equal to the wind velocity in the frame of moving medium. It means that the wave ``stops'' in the frame of a non-moving observer. Consequently, the value of $W_{x}^{*} $ is the maximal possible velocity of the flow in which the wave with frequency $\omega _{0} $ can propagate opposite to the flow direction.

For typical AGW periods of 30--70 min in the upper atmosphere, the values of $\lambda _{x} $ are 580--670 km for waves blocked on the flow, and the phase horizontal velocities contain 710--750 m/s. The frequencies of the waves blocked by the oncoming wind approach the BV frequency (in the reference system of the medium) at different speeds $W_{x} $. Such spectral properties of AGWs are mainly observed from satellites in the polar regions of the thermosphere [Fedorenko et al., 2015].

The experimental dependence of the AGW amplitudes on the wind speed is shown in Fig. 6 according to DE2 satellite measurements [Fedorenko et al., 2018]. For blocked waves, the ratio $A/A_{0} $ should increase linearly with increasing wind speed [Fedorenko et al., 2018], which roughly agrees with observations. In the assumption of blocked waves, we can explain the following main observed properties of AGW in the polar thermosphere [Fedorenko et al., 2015]: 1) the increased wave activity in the region between thermospheric polar vortices; 2) the systematic direction of the AGW propagation towards the wind; 3) the almost linear growth of the wave amplitudes with increasing the wind speed; 4) the predominance of horizontal wavelengths of 500--700 km and the frequencies close to the BV frequency (in the reference system of the medium), which correspond to the condition of blocking AGWs.

\begin{figure}
\centering
\includegraphics[scale=0.8]{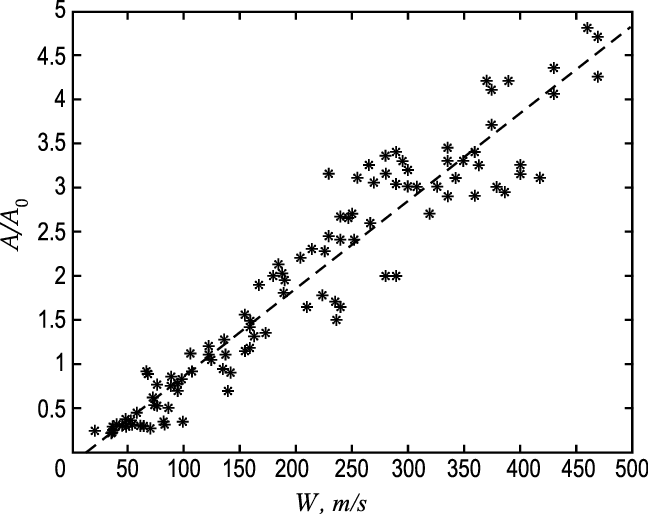}
\caption{Dependence of the AGW amplitude in the polar regions versus the wind speed according to data measured on 6 orbits of DE2 satellite [Fedorenko et al., 2018].}
\end{figure}

\section{DISSIPATION OF ACOUSTIC-GRAVITY WAVES IN THE ATMOSPHERE DUE TO VISCOSITY AND HEAT CONDUCTIVITY}

For analysis of AGW properties in the atmosphere with dissipation, the system of linearized hydrodynamic equations is supplemented with components that take into account viscosity and thermal conductivity [Vadas et al., 2005]. When considering the dissipation of AGWs in the atmosphere in addition to the usually considered velocity gradient, we also take into account the additional components in the Navier-Stokes and transport heat equations, which describe the wave transfer of the momentum and the energy due to the background density gradient [Fedorenko et al., 2020b; Fedorenko et al., 2021]. It gives new local dispersion equation of AGWs in the atmosphere with dissipation. Since the kinematic viscosity coefficient in the atmosphere changes with height, this dispersion equation is valid in the local approximation within thin layers, where the value of $\nu $ can be considered as approximately constant. This is true for waves with vertical scale  $k_{z} >\frac{1}{\nu } \frac{d\nu }{dz} \approx \frac{1}{H} $.

In work [Fedorenko et al., 2021], it was obtained the modified expression for attenuation decrement $\delta$ of AGW. The values of $\delta$ for acoustic and gravitational branches of AGW are shown in Fig. 7 for two values of viscosity coefficient: $\nu =2\cdot 10^{5} $m$^2$/c and $\nu =10^{6} $m$^2$/c. The value of decrement $\delta$ is expected to grow rapidly with an increase of the viscosity coefficient and wih a decrease of the horizontal wavelength.

\begin{figure}[b!]
\centering
\includegraphics[scale=0.8]{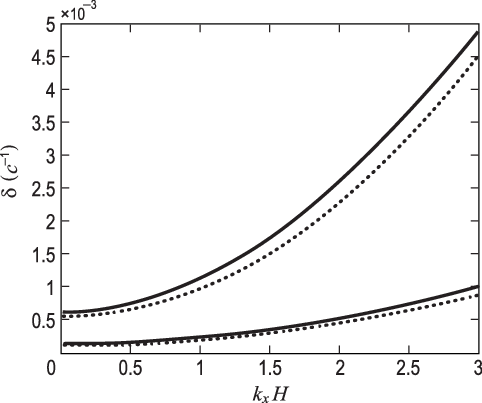}
\caption{Decrement of the attenuation of the gravitational and acoustic branches of AGWs (solid and dotted curves, respectively): $\nu =2\cdot 10^{5} $m$^2$/c (lower curves) and $\nu =10^{6} $m$^2$/c (upper curves), $k_{z} H=1$ [Fedorenko et al., 2021].}
\end{figure}

Attenuation of various types of acoustic-gravity disturbances has certain features, which was studied in [Fedorenko et al., 2020b, Cheremnykh et al. 2021b]. These features are summarized in Table 3. Attenuation decrements for freely propagating AGWs are shown in the two upper rows of the table, and for known evanescent wave modes is below.

Table 3. Attenuation decrements for various types of acoustic-gravity waves in an isothermal dissipative atmosphere [Fedorenko et al., 2020b]. Here, $\nu$ is the kinematic viscosity coefficient, $e=1/3$, $Pr(=0.7)$ is the Prandtl number, and $\alpha=2-\gamma+e(\gamma-1)$.
\begin{center}
\begin{tabular}{|p{1.6in}|p{2.9in}|} \hline
Type of atmospheric disturbance\newline  & Attenuation decrement  \\ \hline
Internal gravity waves\newline  & $\delta _{GR} =\frac{\nu q_{GR}^{2} }{2} \left(1+\frac{1}{\Pr } +\frac{k_{x}^{2} \alpha }{\gamma H^{2} q_{GR}^{2} } \right)$, $q_{GR}^{2} =k_{x}^{2} \frac{N^{2} }{\omega ^{2} } $ \\ \hline
Acoustic waves\newline  & $\delta _{AC} =\frac{\nu q_{AC}^{2} }{2} \left(1+e+\frac{\gamma -1}{\Pr } +\frac{gk_{x}^{2} c_{s}^{2} \alpha }{H\omega ^{4} } \right)$, $q_{AC}^{2} =\frac{\omega ^{2} }{c_{s}^{2} } $ \\ \hline
Non-divergent mode\newline  & $\delta _{ND} =\frac{\nu q_{ND}^{2} }{2} e$, $q_{ND}^{2} =\frac{k_{x} }{H} $ \\ \hline
Inelastic mode  & $\delta _{AE} =\frac{\nu q_{AE}^{2} }{2} \left(\frac{\gamma }{\Pr } +2\right)$, $q_{AE}^{2} =\frac{k_{x} }{H} $  \\ \hline
Vertical BV oscillations  \newline  & $\delta _{BV} =\frac{\nu q_{BV}^{2} }{2} \left(1+e+\frac{\gamma -1}{\Pr } \right)$, $q_{BV}^{2} =\frac{N^{2} }{c_{s}^{2} } $ \\ \hline
Lamb mode & $\delta _{L} =\frac{\nu q_{L}^{2} }{2} \left(1+\frac{\gamma -1}{\Pr } +\frac{ek_{x}^{2} }{q_{L}^{2} } \right)$ , $q_{L}^{2} =k_{x}^{2} +\frac{N^{2} }{c_{s}^{2} } $ \\ \hline
\end{tabular}
\end{center}

\section{EVANESCENT ACOUSTIC-GRAVITY WAVES IN STRATIFIED ATMOSPHERE WITH TAKING INTO ACCOUNT THE EARTH'S ROTATION}

The influence of the Earth's rotation on the spectrum of evanescent waves becomes noticeable in the range of low frequencies that are commensurate with the frequency of the Earth's rotation. The general analysis of the dispersion properties of AGWs in a rotating atmosphere is rather complicated. It is known that the dependence of the vertical component of the Earth's rotation frequency on the horizontal coordinate determines the existence of Rossby waves [Longuet-Higgins, 1964;  Rossby, 1940]. In [Cheremnykh et al., 2022], the high-latitude regions are considered, where it can be taken into account only the vertical component of the Earth's rotation frequency and the waves with a length that is small compared to  the Earth's radius. To consider the wave processes in the Earth's rotating atmosphere, the well-known equations of the ideal gas dynamics were used, the last ones were written in the coordinate system that rotates together with the Earth's atmosphere at a constant angular velocity [Rossby, 1940].

Under a new approach, proposed in [Cheremnykh et al., 2021c], it was obtained many solutions describing evanescent acoustic-gravity waves. These solutions were found by imposing an additional spatial relationship on the components of the perturbed velocity vector of the medium elementary volume. The indicated relationship between the speed components is characterized by the parameter $\alpha $, which can only take real values. It was established that a new spectrum of evanescent acoustic-gravity waves can exist only under the condition 0$<$$\alpha $$<$1, while the previously known spectrum of these waves, modified by taking into account the rotation of the Earth, is realized at arbitrary values of $\alpha $.

The obtained results are conveniently explained by using the diagnostic diagram shown in Fig. 8 (according to [Cheremnykh et al., 2022]). The spectrum of acoustic-gravity waves presented in the Figure consists of acoustic and gravitational regions, as well as two regions of evanescent waves with a continuous spectrum. One region of evanescent waves with frequencies \textbf{$\hat{\omega }_{1} $}, discovered in [Cheremnykh et al., 2021c], is located between the regions of freely propagating acoustic and gravity waves and contains a continuous set of frequencies above the frequency \textbf{$2\Omega $} (Coriolis parameter). The second region of evanescent waves with frequencies \textbf{$\hat{\omega }_{2} $} is realized due to the rotation of the Earth's atmosphere and was discovered for the first time in [Cheremnykh et al., 2022]. This region lies below the frequency \textbf{$2\Omega $}, which is the lower limit of the region of gravitational waves at all wavelengths. The analysis of the obtained solutions showed that these solutions pass into the known evanescent modes at certain values of parameter $\alpha $ [Cheremnykh et al., 2021c].

\begin{figure}
\centering
\includegraphics[scale=0.7]{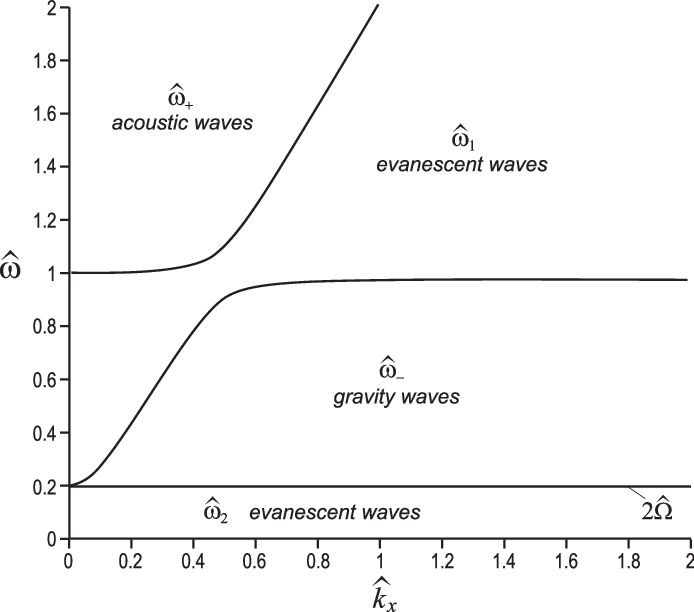}
\caption{Spectral diagram of AGWs with taking into account the effect of the Earth's rotation for high-latitude regions of the atmosphere [Cheremnykh et al., 2022]. Here, $\hat{k}_x=k_xH$ and $\hat{\omega}=\omega/N$.}
\end{figure}

\section{CONCLUSIONS}

The general features of the acoustic-gravity waves propagation in the atmosphere have been studied. The possibility of the existence of new types of evanescent wave modes (the previously unknown $\gamma $-mode and the family of evanescent pseudo-modes) is predicted theoretically. The possibility of observing these modes in the Earth's atmosphere and on the Sun is analyzed [Cheremnykh et al., 2019].

 The possibility of realizing evanescent wave modes on the border of two isothermal media with different temperatures is considered. The peculiarities of matching different type wave solutions on the borders, the polarization of these waves, and the modification of their spectral features are studied. It is analyzed the possibility of applying the results for the clarification of observed features of \textit{f}-mode on the Sun. The theoretical results obtained can be applied in the analysis of observations in the atmosphere at altitudes where there is a sharp change in the temperature, for example, in the lower part of the Earth's thermosphere or in the transition region of the chromosphere-corona on the Sun [Cheremnykh et al., 2019; Fedorenko et al., 2022].

 The effect of vertical temperature inhomogeneity on the propagation of AGW in the Earth's atmosphere is investigated. It is obtained the local dispersion equation of AGWs assuming a slow change in the temperature of the atmosphere with the height. The modification of the acoustic and gravitational regions of freely propagating AGWs depending on the temperature is studied on the spectral plane [Fedorenko et al., 2020a].

  The possibility of realizing \textit{f}- and $\gamma $-modes in the Earth's atmosphere is investigated at the altitudes of local high-altitude extremes of the temperature. An assumption is made regarding the possibility of observing these modes in the atmosphere of the Earth and the Sun. The \textit{f}-mode may be observed near the mesopause with characteristic wavelength $\lambda _{x} \approx 75$km in the Earth's atmosphere, and at the heights of the Sun's chromosphere with characteristic wavelength $\lambda _{x} \approx 1600$km. In the Earth's atmosphere, the $\gamma $-mode may be realized in the regions of maximum temperature, for example, at the height of the stratopause with $\lambda _{x} \approx 100$km [Cheremnykh et al., 2021a].

The propagation of AGW in a spatially inhomogeneous flow is studied. The effect of AGW blocking by oncoming inhomogeneous flow is analyzed. It is established that the spectral properties of the AGWs observed in the polar regions correspond to the waves blocked by the oncoming flow [Fedorenko et al., 2018].

Based on the modified Navier-Stokes and heat transfer equations, it is studied the damping of acoustic-gravity waves. In addition to the usually considered velocity gradient, the modification of these equations consists in taking into account the additional transfer of momentum and energy during the propagation of AGW due to the density gradient. In these assumptions, it is obtained a local dispersion equation of acoustic-gravity waves in an isothermal dissipative atmosphere, as well as an expression for the attenuation decrement. The damping of various types of evanescent acoustic-gravity disturbances is considered. For all considered disturbances, attenuation is expected to increase with increasing wavelength [Fedorenko et al., 2020b; Cheremnykh et al., 2021b; Fedorenko et al., 2021].

It is shown that in the atmosphere rotating with an angular frequency  lower than the frequency $2\Omega $ (Coriolis parameter) the evanescent acoustic-gravity waves  with a continuous spectrum can exist in the atmosphere. On the diagnostic diagram, this spectrum lies below the region of internal gravity waves, and its existence is entirely determined by the rotation of the Earth's atmosphere. It is also shown that the rotation of the atmosphere leads to the modification of the known continuous spectrum of evanescent AGWs with frequencies greater than the Coriolis parameter, on the diagnostic diagram it fills the "forbidden" region between freely propagating acoustic and internal gravity waves. The result is obtained for high-latitude regions of the atmosphere [Cheremnykh et al., 2022].

The work was carried out with the support of the National Research Fund of Ukraine, project 2020.02/0015 "Theoretical and experimental studies of global disturbances of natural and man-made origin in the Earth-atmosphere-ionosphere system".

\end{document}